\documentclass[aps,prl,superscriptaddress,showpacs,twocolumn]{revtex4}
\usepackage{graphicx}
\usepackage{epsfig}
\usepackage{dcolumn}
\usepackage{bm}
\usepackage{psfig}
\usepackage{color}
\newcommand{\BR}{{\cal B}}

\newcommand{\piz}{\pi^0}

\newcommand{\etacp}{\eta_c(2S)}
\newcommand{\psp}{\psi(3686)}
\newcommand{\psip}{\psi(3686)}
\newcommand{\chicJ}{\chi_{cJ}}

\newcommand{\chico}{\chi_{c1}}
\newcommand{\chict}{\chi_{c2}}

\newcommand{\jpsi}{J/\psi}
\newcommand{\EE}{e^+e^-}

\newcommand{\pp}{\pi^+\pi^-}
\newcommand{\kk}{K^+K^-}
\newcommand{\ks}{K^0_S}
\newcommand{\kkp}{K\bar{K}\pi}

\newcommand{\KsKpi}{K^0_SK^\pm\pi^\mp}
\newcommand{\kkpiz}{\kk\piz}

\newcommand{\beq}{\begin{equation}}
\newcommand{\eeq}{\end{equation}}
\newcommand{\bitm}{\begin{itemize}}
\newcommand{\eitm}{\end{itemize}}
\begin{document}

%************************************************************
%\preprint{} \preprint{\vbox{ \hbox{   }
% \hbox{Intended journal: Phys. Rev. Lett.} }}
\title{%%\quad\\[1.0cm]
First observation of the M1 transition $\psp\to \gamma\etacp$}

%\author{AAA}
%\affiliation{Institute of High Energy Physics, Chinese Academy of
%Sciences, Beijing} \collaboration{The BESIII Collaboration}

\author{
%\begin{small}
%\begin{center}
M.~Ablikim$^{1}$, M.~N.~Achasov$^{5}$, D.~J.~Ambrose$^{40}$, F.~F.~An$^{1}$, Q.~An$^{41}$, Z.~H.~An$^{1}$, J.~Z.~Bai$^{1}$, Y.~Ban$^{27}$, J.~Becker$^{2}$, N.~Berger$^{1}$
% ...
\\
\vspace{0.2cm}
(BESIII Collaboration)\\
\vspace{0.2cm} {\it
$^{1}$ Institute of High Energy Physics, Beijing 100049, P. R. China\\
$^{2}$ Bochum Ruhr-University, 44780 Bochum, Germany\\
% ...
$^{5}$ G.I. Budker Institute of Nuclear Physics SB RAS (BINP), Novosibirsk 630090, Russia\\
% ...
$^{27}$ Peking University, Beijing 100871, P. R. China\\
% ...
$^{40}$ University of Rochester, Rochester, New York 14627, USA\\
$^{41}$ University of Science and Technology of China, Hefei 230026, P. R. China\\
% ...
}
%\end{center}
}
\vspace{0.4cm}
%\end{small}

\date{\today}

\begin{abstract}

Using a sample of 106 million $\psp$ events collected with the
BESIII detector at the BEPCII storage ring, we have made the first
measurement of the M1 transition between the radially excited charmonium $S$-wave
spin-triplet and the radially excited $S$-wave
spin-singlet states: $\psip\to\gamma\etacp$. Analyses of the processes $\psp\to
\gamma\etacp$ with $\etacp\to \KsKpi$ and $\kkpiz$ gave an
$\etacp$ signal with a statistical significance of greater than 10
standard deviations under a wide range of assumptions about the
signal and background properties.  The data are used to obtain
measurements of the $\etacp$ mass ($M(\etacp)=3637.6\pm
2.9_\mathrm{stat}\pm 1.6_\mathrm{sys}$~MeV/$c^2$), width
($\Gamma(\etacp)=16.9\pm 6.4_\mathrm{stat}\pm
4.8_\mathrm{sys}$~MeV), and the product branching fraction ($\BR(\psp\to \gamma\etacp)\times \BR(\etacp\to
K\bar K\pi) = (1.30\pm 0.20_\mathrm{stat}\pm
0.30_\mathrm{sys})\times 10^{-5}$). Combining our result with a
BaBar measurement of $\BR(\etacp\to K\bar K \pi)$, we find the
branching fraction of the M1 transition to be
$\BR(\psp\to\gamma\etacp) = (6.8\pm 1.1_\mathrm{stat}\pm
4.5_\mathrm{sys})\times 10^{-4}$.

\end{abstract}

\pacs{13.25.Gv, 13.20.Gd, 14.40.Pq}

\maketitle

%%%%%%%%%%%%%%%%%%%%%%%%%%%%%%%%%%%%%%%%%%%%%%%%%%%%%%%%%%%%%%%%
%%%%%     Introduction       Part                  %%%%%%%%%%%%%
%%%%%%%%%%%%%%%%%%%%%%%%%%%%%%%%%%%%%%%%%%%%%%%%%%%%%%%%%%%%%%%%
Recent discoveries of charmonium and charmonium-like states
above the open-charm production threshold have generated great
interest. Intensive efforts to incorporate these states into the quark-model picture of
hadrons have led to the development of models to explain some or
all of the new states~\cite{many}.  The charmonium states below
the open-charm production threshold are relatively well understood, with
the notable exception of the spin singlets. These include the $P$-wave
state $h_c$ and the $S$-wave ground state $\eta_c$ and its first radial
excitation $\etacp$~\cite{PDG2010}.  These are
experimentally challenging because of the low production rates and
spin-parity quantum numbers that are inaccessible in direct $\EE$ annihilations.

The $\etacp$ was first observed by the Belle collaboration in the
process $B^\pm\to K^\pm \etacp$, $\etacp\to K_S^0K^\pm
\pi^\mp$~\cite{babar_B2Ketacp}.  It was confirmed in the
two-photon production of $\KsKpi$~\cite{babar_2gamEtacp,
cleo_2gamEtacp}, and in the double-charmonium production process
$e^+e^-\to J/\psi c\bar{c}$~\cite{belle_Jpsiccbar,
babar_Jpsiccbar}. Combining the world-average
values~\cite{PDG2010} with the most recent results from Belle and
BaBar on two-photon fusion into hadronic final states other than
$\KsKpi$~\cite{etacp_2gam_Belle_new, etacp_2gam_Babar_new}, one
obtains updated averages of the $\etacp$ mass and width of
$3637.7\pm 1.3~{\rm MeV}/c^2$ and $10.4\pm 4.2~{\rm MeV}$,
respectively.

The production of the $\etacp$ through a radiative transition from the
$\psp$ requires a charmed-quark spin-flip and,
thus, proceeds via a magnetic dipole (M1) transition. The branching
fraction has been calculated by many authors, with predictions in
the range $\BR(\psp\to \gamma \etacp)= (0.1-6.2)\times
10^{-4}$~\cite{gaoky}. A recent calculation~\cite{zhaoq} that
includes contributions from loops containing meson pairs finds
a strong cancellation that results in a partial width of $(0.08\pm
0.03)$~keV and a branching fraction of $(2.6\pm 1.0)\times
10^{-4}$; while a calculation using the light-front quark model
and a 2S state harmonic oscillator wave function to present the 2S charmonium state gives a
transition rate of $3.9\times 10^{-4}$~\cite{mabq}.
Experimentally, this transition has been searched for by Crystal
Ball~\cite{cbal}, BES~\cite{ycz}, CLEO~\cite{etacp_cleo_c}
%with 26 million $\psp$ events,
and most recently by BESIII through
$\etacp\to VV$ \cite{etacp_VV}. No convincing signal was observed
in any of these searches.

%%%%%%%%%%%%%%%%%%%%%%%%%%%%%%%%%%%%%%%%%%%%%%%%%%%%%%%%%%%%%%%%
%%%%%  Data sample and MC generator                %%%%%%%%%%%%%
%%%%%%%%%%%%%%%%%%%%%%%%%%%%%%%%%%%%%%%%%%%%%%%%%%%%%%%%%%%%%%%%

In this Letter, we report the first observation of $\psp\to \gamma\etacp$, with $\etacp\to
\KsKpi$ and $K^+K^-\pi^0$. The data sample for this analysis
consists of an integrated luminosity of 156~pb$^{-1}$ ($106$
million events) produced at the peak of the $\psp$
resonance~\cite{npsp} and collected in the BESIII detector~\cite{bes3}. An additional 42~pb$^{-1}$ of data were
collected at a center-of-mass energy of $\sqrt{s}$=3.65~GeV to
determine non-resonant continuum background contributions.

The BESIII detector, described in detail in Ref.~\cite{bes3}, has
an effective  geometrical acceptance of 93\% of $4\pi$. A
small-cell, helium-based main drift chamber (MDC) in a 1-T
magnetic field provides a charged-particle momentum resolution of
0.5\% at 1~GeV/$c$, and specific-ionization ($dE/dx$) measurements
for particle identification with a resolution better than 6\% for
electrons from Bhabha scattering. The cesium iodide
electromagnetic calorimeter (EMC) measures photon energies with
resolutions at 1.0~GeV of 2.5\% and 5\% in the detector's barrel
($|{\rm cos} \theta| < 0.8$, where $\theta$ is the polar angle
with respect to the $e^+$ direction) and endcaps ($0.86 < |{\rm
cos} \theta|< 0.92$) regions, respectively.  Additional particle
identification is provided by a time-of-flight system (TOF) with a
time resolution of 80~ps (110~ps) for the barrel (endcaps).

%
%I see no usefulness in providing information about the muon detector,
%which is not used in this analysis, and is described in Ref. 16.
%The muon system provides a $\sim
%2$ cm position resolution and identifies muons with momenta
%larger than 0.5~GeV/$c$.

%%%%%%%%%%%%%%%%%%%%%%%%%%%%%%%%%%%%%%%%%%%%%%%%%%%%%%%%%%%%%%%%
%%%%%     Event Selection                          %%%%%%%%%%%%%
%%%%%%%%%%%%%%%%%%%%%%%%%%%%%%%%%%%%%%%%%%%%%%%%%%%%%%%%%%%%%%%%

Reconstructed charged tracks other than daughters of $K^0_S$
candidates are required to pass within 1~cm of the $e^+e^-$
annihilation interaction point (IP) transverse to the beam line
and within 10~cm of the IP along the beam axis. Each track is
required to have a good-quality fit and to satisfy the condition
$|{\rm cos}\theta|<0.93$.  Charged-particle identification (PID)
is based on combining the $dE/dx$ and TOF information
to construct a $\chi^{2}_{{\rm PID}}(i)$.
%in the
%variable $\chi^{2}_{{\rm PID}}(i)= (\frac{dE/dx_{\rm
%measured}-dE/dx_{\rm expected}}{\sigma_{dE/dx}})^2+
%(\frac{TOF_{\rm measured}-TOF_{\rm expected}}{\sigma_{TOF}})^2$.
The values $\chi^{2}_{{\rm PID}}(i)$ and the corresponding
confidence levels ${\rm Prob_{PID}}(i)$ are calculated for each
charged track for each particle hypothesis $i$ (pion, kaon or
proton).

A neutral cluster in the EMC must satisfy fiducial and
shower-quality requirements to be accepted as a good photon
candidate. Showers must have a minimum energy of 25~MeV and be
detected in either the barrel or endcap regions, as previously
defined.  EMC timing requirements are used to suppress noise and
energy deposits unrelated to the event.

In selecting $\gamma \KsKpi$ ($\gamma\kkpiz$) events, the decay
signal $K_S^0\to \pi^+\pi^-$ ($\piz\to\gamma\gamma$) is used to
tag the $K_S^0$ ($\piz$). Candidate events must therefore have
exactly four (two) charged tracks with zero net charge and at
least one (three) good photon(s) for the $\gamma \KsKpi$
($\gamma\kkpiz$) decay mode.

$K^0_S$ candidates are selected with secondary-vertex fits to all
pairs of oppositely charged tracks in the event, assuming pion
masses. The combination with the best fit quality is chosen and
the event is kept for further analysis if the invariant mass is
within 7~MeV/$c^2$ of the expected $\ks$ mass, and the secondary
vertex is at least 0.5~cm from the interaction point. To suppress
$\gamma K^0_S K^0_S$ events, the remaining tracks are required not
to form a good $\ks$ candidate. The fitted $K^0_S$ information is
used as input for the subsequent kinematic fit of the complete
event.

The $\gamma\KsKpi$ candidates are then subjected to a
four-constraint ($4C$) kinematic fit, with the constraints
provided by four-momentum conservation.  The discrimination of
charge-conjugate channels ($K_S^0K^+\pi^-$ or $K_S^0K^-\pi^+$) and
the selection of the best photon among multiple candidates are
achieved by minimizing $\chi^2 = \chi_{4C}^2 + \chi^2_{{\rm
PID}}(K) + \chi^2_{{\rm PID}}(\pi)$, where $\chi_{4C}^2$ is the
chi-square of the $4C$ kinematic fit. Events with $\chi_{4C}^2<50$
are accepted as $\gamma\KsKpi$ candidates. For $\gamma\kkpiz$
candidates, both charged tracks must satisfy the criterion that
the kaon-hypothesis probability ${\rm Prob_{PID}}(K)$ is larger
than both 0.001 and the probability of any other hypothesis.  A
five-constraint ($5C$) kinematic fit, with the $\pi^0$ mass as the
additional constraint, is used to select the best transition
photon and the $\piz\to \gamma\gamma$ combination. Events with
$\chi_{5C}^2<30$ are accepted as $\gamma\kkpiz$ candidates.

%%%%%%%%%%%%%%%%%%%%%%%%%%%%%%%%%%%%%%%%%%%%%%%%%%%%%%%%%%%%%%%%
%%%%%     Backgrounds                              %%%%%%%%%%%%%
%%%%%%%%%%%%%%%%%%%%%%%%%%%%%%%%%%%%%%%%%%%%%%%%%%%%%%%%%%%%%%%%

We use the program {\sc lundcrm}~\cite{lundcrm} to generate
inclusive Monte Carlo (MC) events for background studies. The
signal is generated with the expected angular distribution for
$\psp\to \gamma \etacp$, and the subsequent $\etacp\to \KsKpi$ and
$\kkpiz$ decays are generated according to phase space. The
detector response is simulated with a {\sc geant4}-based
package~\cite{geant4} that has been tuned to match the performance
of the detector components.

The $\psp\to \gamma\etacp$, $\etacp\to \KsKpi$ $(\kkpiz)$ signal
suffers significantly from background contributions from leptonic
$J/\psi$ decays and $\jpsi\to\kk$ in $\psp \to \pi^+\pi^-J/\psi$,
and $\psp \to \eta J/\psi$ with $\eta\to \pi^+\pi^-\pi^0$
$(\gamma\gamma)$. For the $\gamma\KsKpi$ channel, these background
contributions are suppressed by requiring that the recoil mass of
all $\pp$ pairs be less than $3.05$~GeV/$c^2$. For the
$\gamma\kkpiz$ channel, this type of contamination is removed by
requiring that the invariant mass of the two charged tracks, assuming they are muons, to be
less than $2.9$~GeV/$c^2$. The remaining
dominant background sources are (1) $\psp\to \KsKpi$ ($\kkpiz$)
events with a fake photon candidate; (2) events with the same
final states including $\KsKpi\gamma_{{\rm ISR}/{\rm FSR}}$
($\kkpiz\gamma_{{\rm ISR}/{\rm FSR}}$) with the photon from
initial- or final-state radiation (\mbox{ISR}, \mbox{FSR}) and
$\psp\to\omega\kk$ with $\omega\to\gamma\piz$; and (3) events
with an extra photon, primarily from $\psp\to \pi^0\KsKpi$
($\piz\kkpiz$) with $\piz\to \gamma\gamma$.  MC studies
demonstrate that contributions from all other known processes are negligible.

The events in the first category, with a fake photon incorporated
into the kinematic fit, produce a peak in the $\KsKpi$ ($\kkpiz$)
mass spectrum close to the expected $\etacp$ mass, with a sharp
cutoff due to the 25-MeV photon-energy threshold.

Because the fake
photon adds no information to the fit, its inclusion distorts the
mass measurement. We therefore determine the mass from a modified
kinematic fit in which the the magnitude
of the photon momentum is allowed to freely float ($3C$ for $\gamma\KsKpi$ and $4C$ for
$\gamma\kkpiz$). In the case of a fake photon, the momentum
tends to zero, which improves the background separation
with minimal distortion of the signal line shape~\cite{etacp_VV}.

Background contributions from $\psp\to\KsKpi$ $(\kkpiz)$ and
$\psp\to\KsKpi\gamma_{\rm FSR}$ $(\kkpiz\gamma_{\rm FSR})$ are
estimated with MC distributions for those processes
normalized according to a previous measurement of the branching ratios~\cite{psipToK*K}. \mbox{FSR} is
simulated in our MC with {\sc photos}~\cite{photos}, and the \mbox{FSR}
contribution is scaled by the ratio of the \mbox{FSR} fractions in data
and MC for a control sample of $\psp\to \gamma \chi_{c{J}}$
(${J}=0$ or $1$) events.  For this study the $\chi_{c{J}}$ is
selected in three final states with or without an extra \mbox{FSR}
photon, namely $\KsKpi(\gamma_{\rm FSR})$,
$\pi^+\pi^-\pi^+\pi^-(\gamma_{\rm FSR})$, and $\pi^+\pi^-K^+K^-
(\gamma_{\rm FSR})$, as described in Ref.~\cite{etacp_VV}.
Background contributions from the continuum process
$e^+e^-\to\gamma^*\to \KsKpi(\gamma_{\rm FSR})$
$(\kkpiz(\gamma_{\rm FSR}))$ and the \mbox{ISR} process $e^+e^-\to
\gamma^*\gamma_{\rm ISR}\to \KsKpi\gamma_{\rm
ISR}~(\kkpiz\gamma_{\rm ISR})$ are estimated with data collected
at $\sqrt{s}=3.65$~GeV corrected for differences in the
integrated luminosity and the cross section, and with particle momenta and energies scaled to account for the beam-energy
difference. MC simulations show that the $\KsKpi$ ($\kkpiz$) mass
spectra are similar for \mbox{FSR} and \mbox{ISR} events. Events without
radiation have the same mass distribution independently of originating from a resonant $\psp$ decay or from the
non-resonant continuum production. Thus, the background shapes from
$\KsKpi~(\kkpiz)$ and $\KsKpi\gamma_{\rm
ISR/FSR}~(\kkpiz\gamma_{\rm ISR/FSR})$ are described by the sum of
the MC-simulated $\KsKpi~(\kkpiz)$ and $\KsKpi\gamma_{\rm FSR}
~(\kkpiz\gamma_{\rm FSR})$ invariant-mass shapes, with the
proportions fixed according to the procedure described above. The
shapes of backgroung mass distributions from $\psp\to\omega\kk$ with
$\omega\to\gamma\piz$ are parameterized with a double-Gaussian
function, and its level is measured with the same data sample and fixed
in the final fit.

The third type of background, that with an extra photon,
$\pi^0\KsKpi~(\pi^0\kkpiz)$, is measured with data and normalized
according to the simulated contamination rate. It contributes a
smooth component around the $\chi_{cJ}$ $(J=1,2)$ mass region with
a small tail in the $\etacp$ signal region that is described
by a Novosibirsk function~\cite{NovosibirskFun} (Gaussian
function) for the $\piz\KsKpi$ ($\piz\kkpiz$) background. The shape and size of
%parameters for the functions and the corresponding numbers of
%events are
this background is
fixed in the fit.

%%%%%%%%%%%%%%%%%%%%%%%%%%%%%%%%%%%%%%%%%%%%%%%%%%%%%%%%%%%%%%%%
%%%%%        mass spectrum and FIT                 %%%%%%%%%%%%%
%%%%%%%%%%%%%%%%%%%%%%%%%%%%%%%%%%%%%%%%%%%%%%%%%%%%%%%%%%%%%%%%

\begin{figure*}[htb]
\centering
  \includegraphics[width=0.49\textwidth]{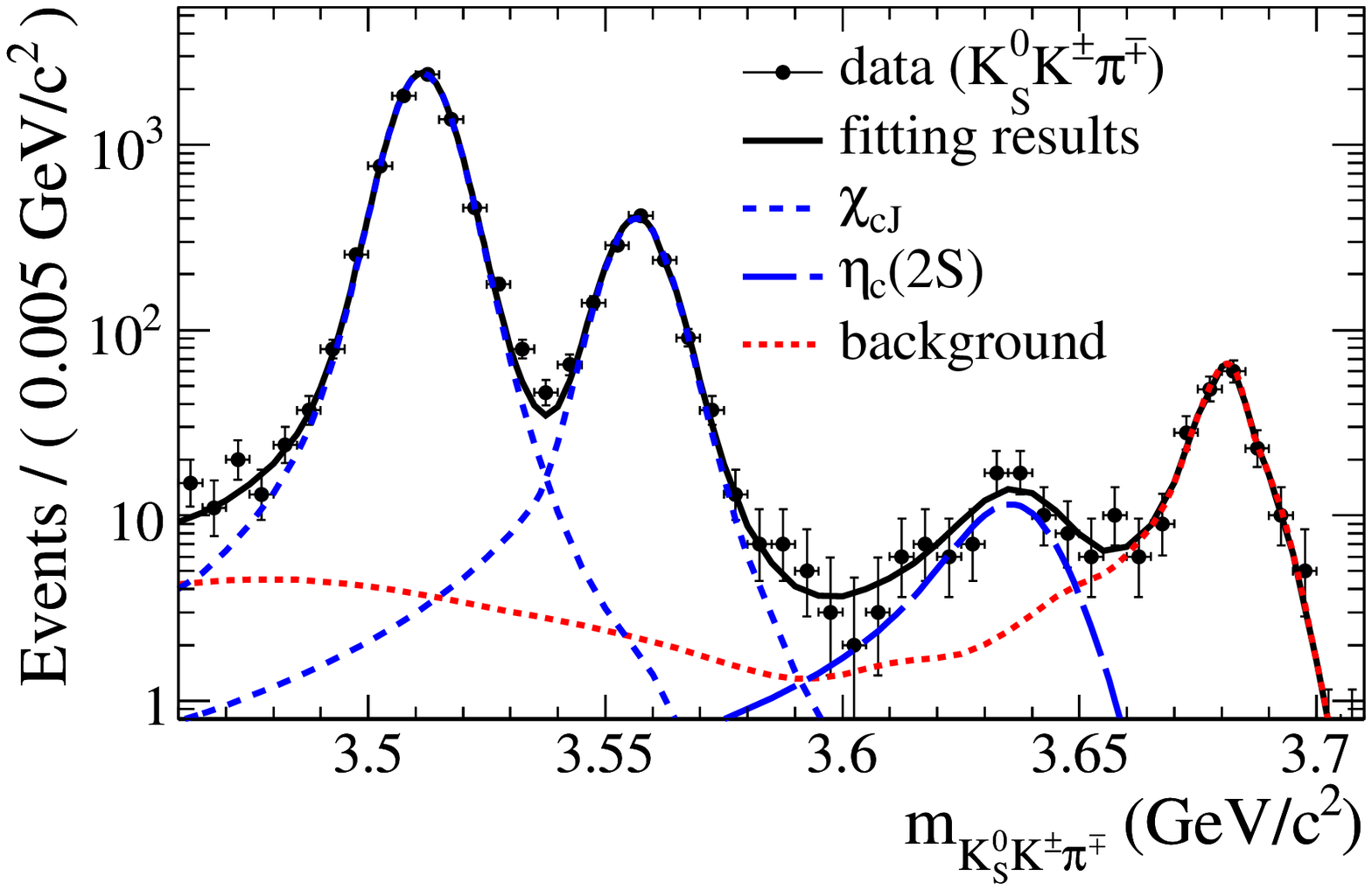}
  \includegraphics[width=0.49\textwidth]{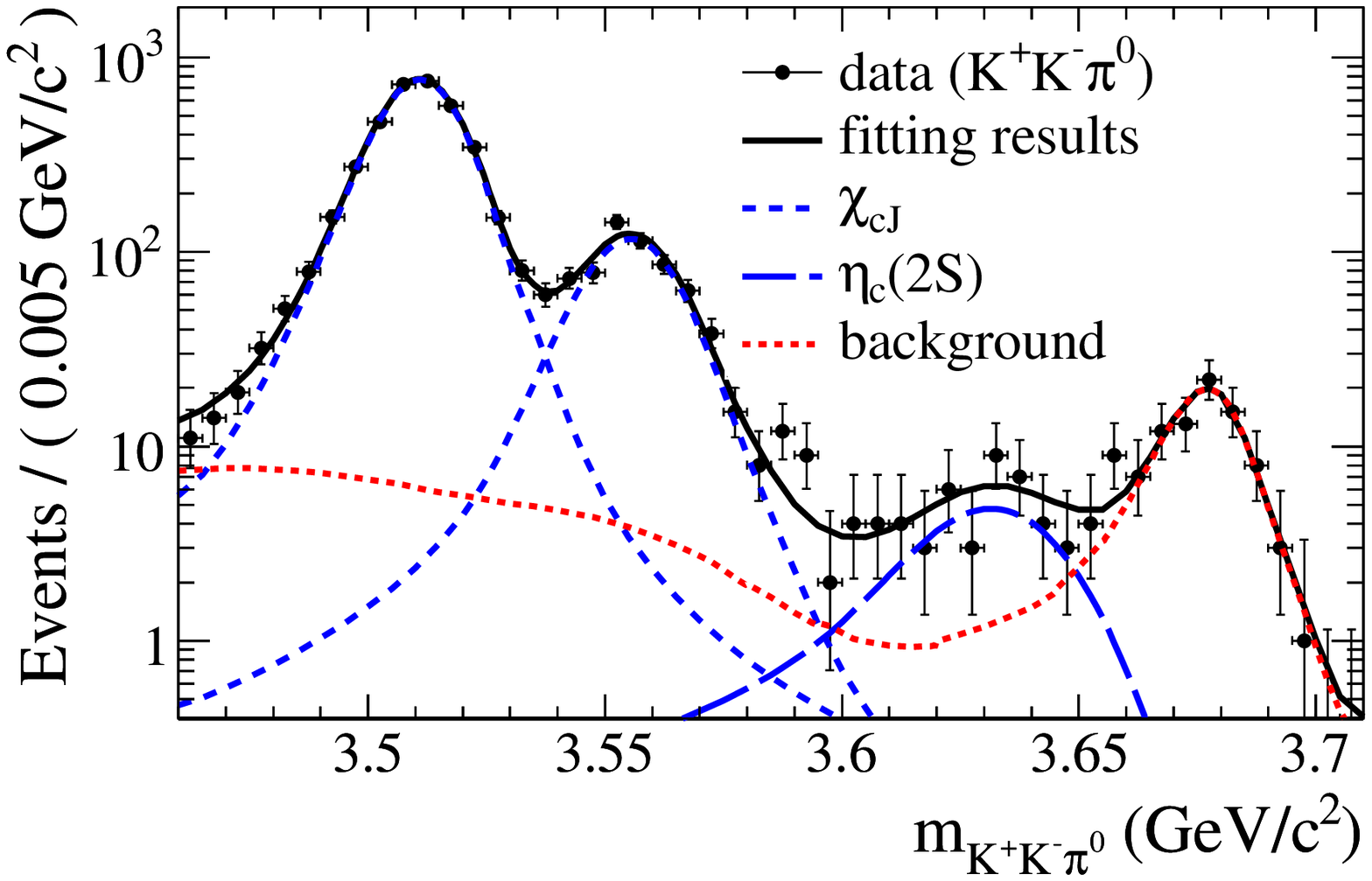}
%  \put(-120, -7){$\rm{m_{\kkpiz}}~(\rm{GeV}/c^2)$}
%  \put(-355, -7){$\rm{m_{\KsKpi}}~(\rm{GeV}/c^2)$}
\caption{The invariant-mass spectrum for $\KsKpi$ (left panel),
$\kkpiz$ (right panel), and the simultaneous likelihood fit to the
three resonances and combined background sources as described in
the text.} \label{pic_fit_etacp}
\end{figure*}

The mass spectra for the $\KsKpi$ and $\kkpiz$ channels are fitted
%separately and
simultaneously to extract the yield, mass and width
of $\etacp$. To better determine the background and mass
resolution from the data, the mass spectra are fitted over a range
(3.46-3.71~GeV/$c^2$) that includes the $\chico$ and $\chict$
resonances as well as the $\etacp$ signal. The final mass spectra
and the likelihood fit results are shown in
Fig.~\ref{pic_fit_etacp}. Each fitting function includes four
components, namely the $\etacp$, $\chico$, $\chict$, and the
summed background described above.  Line shapes for the $\chico$
and $\chict$ are obtained from MC simulations and convolved with
Gaussian functions to accommodate for the mass-scale and
resolution differences from data. For both modes, the $\chico$ and
$\chict$ widths are fixed to the PDG values~\cite{PDG2010}. Based
on MC studies, the mass shift and resolution for the resonances
are found to vary linearly as a function of the $\KsKpi~(\kkpiz)$
invariant mass. These parameters are extrapolated from the
$\chico$ and $\chict$ to the $\etacp$.

The line shape for the $\etacp$ produced in the M1 transition of
the $\psp$ is assumed to have the form $(E^3_{\gamma}\times
BW(m)\times f_d(E_{\gamma}) \times\epsilon(m))\otimes G(\delta
m,\sigma$), where $m$ is the invariant mass of the $\KsKpi$ or
$\kkpiz$, $E_{\gamma} = (m^2_{\psp}-m^2)/2m_{\psp}$ is the energy
of the transition photon in the rest frame of $\psp$, $BW(m)$ is
the Breit-Wigner function for $\etacp$, $f_d(E_{\gamma})$ is a
function that damps the diverging tail originating from the $E^3_{\gamma}$
dependence, $\epsilon(m)$ is the mass-dependent efficiency
function determined by a full simulation of the signal, and
$G(\delta m,\sigma)$ is a Gaussian function describing the mass
shift and the detector resolution. For the damping function we use
a functional form introduced by the KEDR
collaboration~\cite{KEDR_Jpsi2gEtac}: $f_d=E^2_0 / [E_{\gamma}E_0
+(E_{\gamma}-E_0)]^2$, where $E_0 = (m^2_{\psp} -
m_{\etacp}^2)/{2m_{\psp}}$ is the peak energy of the transition
photon. To assess the sensitivity of our results to the choice of
this function, we also consider an alternative form used by
CLEO~\cite{CLEO_Jpsi2gEtac}: $f_d=\exp(-\frac{E^2_\gamma}
{8\beta^2})$, with CLEO's fitted value of $\beta=(65.0\pm
2.5)$~MeV.

The fit shown in Fig.~\ref{pic_fit_etacp} has a $\chi^2$ of 72 for
79 degrees of freedom. The results for the yields of $\etacp$
events are $81\pm14$ for the $\KsKpi$ channel and $46\pm11$ for
the $\kkpiz$ channel. Consistent yields are found for separate %and simultaneous fits
fits to the two channels~\cite{sep_fit}. The $\KsKpi$ channel
determines primarily the precision for the $\etacp$ mass and width
measurements in the simultaneous fit with the results
$M_{\etacp}=3637.6\pm 2.9$~MeV/$c^2$ and $\Gamma_{\etacp} =
16.9\pm 6.4~{\rm MeV}$, respectively. The combined statistical
significance of the signal in the two modes is $11.1\sigma$, which
is obtained by comparing the likelihoods of the fits with and
without the $\etacp$ signal. The robustness of this result was
tested by considering variations of the resonant line shapes,
background assumptions and other systematic effects. In all the
cases, the statistical significance is found to be larger than
$10.2\sigma$.

%The individual fit of the $\KsKpi$ spectrum gives
%$M_{\etacp({\KsKpi})} = (3637.0\pm 2.7)~\mathrm{MeV}/c^2$,
%$\Gamma_{\etacp({\KsKpi})} = (14.4\pm 5.3)~\mathrm{MeV}$,
%$N_{\etacp({\KsKpi})} = 77\pm 13$ with a purely statistical
%significance of $9.5\sigma$; and the fit of the $\kkpiz$
%distribution gives $M_{\etacp({\kkpiz})} = (3656\pm
%31)~\mathrm{MeV}/c^2$, $\Gamma_{\etacp({\kkpiz})} = (60\pm
%39)~\mathrm{MeV}$, $N_{\etacp({\kkpiz})} = 72\pm 32$ with a purely
%statistical significance of $5.7\sigma$.

Combining the observed number of signal events with the
efficiencies of $25.6\%$ and $20.2\%$ for the $\KsKpi$ and
$\kkpiz$ final states, respectively, from full simulations of the
signal with the measured $\etacp$ mass and width, we find the
product branching fractions $\BR(\psp\to \gamma\etacp)\times
\BR(\etacp\to \KsKpi) = (4.31\pm 0.75)\times 10^{-6}$, and
$\BR(\psp\to \gamma\etacp)\times \BR(\etacp\to\kkpiz) = (2.17\pm
0.52)\times 10^{-6}$, where the errors are statistical only. The
ratio of the branching fractions agrees well with the isospin
symmetry expectation of 2:1 between $\KsKpi$ and $\kkpiz$.  The
product branching fraction for $\psp\to \gamma\etacp$, $\etacp\to
\kkp$ can be obtained by doubling the sum of the $\KsKpi$ and
$\kkpiz$ branching fractions to obtain $\BR(\psp\to \gamma\etacp)
\times \BR(\etacp\to \kkp) = (1.30\pm 0.20)\times 10^{-5}$, where
the error takes into account the correlation between the two
measured branching fractions from the simultaneous fit.

The systematic uncertainties in the branching fraction, $\etacp$
mass and $\etacp$ width measurements are summarized in
Table~\ref{tab_sys}.
The uncertainties due to the choice of the background shape, the
damping function, the fitting range and the linear extrapolated mass shift for $\etacp$ are common among the three
measurements and are determined together. The systematic errors in
the mass and width due to the $\KsKpi(\gamma_{\rm ISR/FSR})$
($\kkpiz(\gamma_{\rm ISR/FSR})$) background shape are evaluated by
changing the relative ratio of the $\KsKpi$ ($\kkpiz$) background
events with and without radiation. The uncertainties from $\piz
\KsKpi$ ($\piz \kkpiz$) background shape are estimated by changing
the function parameterizing the measured mass spectrum. The
uncertainty due to the choice of damping function is estimated from the
difference between results obtained with the default (KEDR) and
alternative (CLEO) functional forms.  The uncertainties due to the
choice of fitting range are estimated by taking the largest
differences between results found with the standard fitting range
and those obtained using alternative ranges. The uncertainties from the linear extrapolation of the mass shifts from $\chico$ and $\chict$ to $\etacp$ are estimated from the maximum changes in the fitting results obtained by varying the mass shifts within their errors.

\begin{table}
\centering \caption{The absolute systematic uncertainties in the $\etacp$
mass (in MeV/$c^2$), width (in MeV) and the relative systematic error (in \%) in $\BR\BR$, the product branching
fraction $\BR(\psp\to \gamma\etacp)\times \BR(\etacp\to\kkp)$, measurements.} \label{tab_sys}
  \bigskip
  \begin{tabular}{lccc}
    \hline
    Source                      & Mass    & Width  &  $\BR\BR$ \\
    \hline
    \hline
    Background shape            & 1.3     & 2.6       &  9.9 \\
    Damping function            & 0.7     & 4.0       & 19.6 \\
    Fitting range               & 0.1     & 0.4       &  1.3 \\
    Mass shift                  & 0.6     & 0.2       &  0.4 \\
    \hline
    Tracking                    & -       & -         & 4.0 \\
    Photon reconstruction       & -       & -         & 1.3 \\
    Particle identification     & -       & -         & 1.3 \\
    $\ks$ reconstruction        & -       & -         & 2.3 \\
    Kinematic fitting           & -       & -         & 3.9 \\
    $\etacp$ decay dynamics     & -       & -         & 1.5 \\
    Number of $\psp$ events     & -       & -         & 4.0 \\
    \hline
    Total                       & 1.6     & 4.8       & 23.3\\
    \hline
  \end{tabular}
\end{table}

The branching fraction measurement is affected by additional
effects that enter through the yield determination, including
those associated with charged-particle tracking, photon
reconstruction, particle identification, $K_S^0$ reconstruction,
and kinematic fitting ($\chi^2$ requirement), all of which are
estimated with control samples in the data~\cite{gammaV}. The
effect of the uncertainty in the dynamics of the decay $\etacp\to
\KsKpi~(\kkpiz)$, which is treated as phase space in our default
signal MC, is estimated with an alternative MC replicating the
Dalitz distribution of $\etacp\to \KsKpi$ decay recently measured
by the Belle collaboration~\cite{Belle_B2KKsKpi}. A 0.8\% (3.0\%)
relative difference in the efficiency was found between the
default and alternative MC samples for $\KsKpi$ ($\kkpiz$),
leading to a 1.5\% difference in the total branching ratio, which
we take as a systematic error.  Finally, there is an overall 4\%
uncertainty in the branching fraction associated with the
determination of the total number of $\psip$ events in our data
sample~\cite{npsp}.

We assume that all the sources of systematic uncertainties are
independent and combine them in quadrature to obtain the
overall systematic uncertainties given in Table~\ref{tab_sys}.
The total systematic uncertainties on the mass and width
measurements are 1.6~MeV/$c^2$ and 4.8~MeV, respectively; the
total relative systematic uncertainty on the product branching fraction
$\BR(\psp\to\gamma\etacp)\times \BR(\etacp\to K\bar K\pi)$ is
$23.3\%$. Using the measurement of $\BR(\etacp\to
K\bar{K}\pi)=(1.9\pm 0.4\pm 1.1)\%$ from the BaBar
experiment~\cite{BToXK_Babar}, we find an M1-transition branching
fraction of $\BR(\psp\to \gamma\etacp) = (6.8\pm 1.1\pm 4.5)\times
10^{-4}$, where the systematic error is dominated by that of the
BaBar result.

%%%%%%%%%%%%%%%%%%%%%%%%%%%%%%%%%%%%%%%%%%%%%%%%%%%%%%%%%%%%%%%%
%%%%%     Summary       Part                       %%%%%%%%%%%%%
%%%%%%%%%%%%%%%%%%%%%%%%%%%%%%%%%%%%%%%%%%%%%%%%%%%%%%%%%%%%%%%%

In summary, we report the first observation of the M1
transition $\psp\to \gamma \etacp$ through the decay processes
$\psp\to \gamma\KsKpi$ and $\gamma\kkpiz$. We measure the mass of
the $\etacp$ to be $3637.6\pm 2.9\pm 1.6$~MeV/$c^2$, the width
$16.9\pm 6.4\pm 4.8$~MeV, and the product branching fractions
$\BR(\psp\to \gamma\etacp)\times \BR(\etacp\to \kkp) = (1.30\pm
0.20\pm 0.30)\times 10^{-5}$, where the quoted uncertainties are
statistical and systematic, respectively.  The main systematic
limitations to these measurements arise from the choice of the
functional form for the damping factor in the $\etacp$ line shape
and from uncertainty in the choice of the background line shapes.
Our results are consistent with previously published values and
limits, and the branching-fraction measurement of the M1
transition $\psp\to \gamma\etacp$ of $(6.8\pm 1.1\pm 4.5)\times
10^{-4}$ agrees with theoretical calculations and naive estimates
based on the $\jpsi\to \gamma \eta_c$
transition~\cite{etacp_cleo_c}.

%%%%%%%%%%%%%%%%%%%%%%%%%%%%%%%%%%%%%%%%%%%%%%%%%%%%%%%%%%%%%%%%
%%%%%    acknowledgments       Part                %%%%%%%%%%%%%
%%%%%%%%%%%%%%%%%%%%%%%%%%%%%%%%%%%%%%%%%%%%%%%%%%%%%%%%%%%%%%%%

We would like to thank S.~Eidelman and A.~Vinokurova for supplying
the details of the Dalitz plot of $\etacp\to \KsKpi$ decay from
the Belle experiment.
The \mbox{BESIII} collaboration thanks the staff of BEPCII and the computing center for their hard efforts. This work is supported in part by the Ministry of Science and Technology of China under Contract No. 2009CB825200; National Natural Science Foundation of China (NSFC) under Contracts Nos. 10625524, 10821063, 10825524, 10835001, 10935007, 11125525; Joint Funds of the National Natural Science Foundation of China under Contracts Nos. 11079008, 11179007; the Chinese Academy of Sciences (CAS) Large-Scale Scientific Facility Program; CAS under Contracts Nos. KJCX2-YW-N29, KJCX2-YW-N45; 100 Talents Program of CAS; Istituto Nazionale di Fisica Nucleare, Italy; U. S. Department of Energy under Contracts Nos. DE-FG02-04ER41291, DE-FG02-91ER40682, DE-FG02-94ER40823; U.S. National Science Foundation; University of Groningen (RuG) and the Helmholtzzentrum fuer Schwerionenforschung GmbH (GSI), Darmstadt; WCU Program of National Research Foundation of Korea under Contract No. R32-2008-000-10155-0


\begin{thebibliography}{99}
\bibitem{NovosibirskFun} The Novosibirsk function is defined
as $f(m)=A_S\exp(-0.5\{\ln^2[1+\Lambda\tau(m-m_0)]/
\tau^2+\tau^2\})$, where $\Lambda = \sinh(\tau\sqrt{\ln4})/
(\sigma\tau\sqrt{\ln4})$, the peak position is $m_0$, the width is
$\sigma$, and $\tau$ is the tail parameter.
\end{thebibliography}
\end{document}